\documentstyle[prb,multicol,aps]{revtex}
\draft

\begin{document}

\title{Dynamic percolation and Slow Relaxation in Glass-like Materials}
\author{Alexei V\'azquez and Oscar Sotolongo-Costa}
\address{Department of Theoretical Physics. Havana University. Havana
10400, Cuba.}
\date{\today}
\maketitle

\begin{abstract}
Glass-like materials are nonequilibrium systems where the relaxation
time may exceed reasonable time scales of observations. In the present
paper a dynamic percolation model is introduced in order to explain the
principal properties of glass-like materials near the dynamic
transition. Here, contrary to conventional percolation problems,
clusters are groups of particles dynamically correlated.  Introducing a
size dependent relaxation time and the scaling hypothesis for the
distribution of dynamically correlated clusters the two step relaxation
predicted by the mode coupling theory and observed in experiments is
obtained.
\end{abstract}

\pacs{61.20.Lc}

\begin{multicols}{2}

\section{Introduction}

In common second order transitions the new phase has as distinguish
property the existence of long range order, leading to macroscopic
structures like ferromagnetic domains. However, while it is believed
that glass-like transitions are second order transitions, the
glass-like state does not exhibit the conventional long range order but
a short range order, at a mesoscopic scale, is preferred. The
relaxation time of these systems becomes as long as, or longer than,
the observation time and, therefore, the system cannot reach
equilibrium within observation time. Thus, it is not clear if the
glass-like transition is a true thermodynamic phase transition or a
slow relaxation phenomena, usually referred to as dynamic or freezing
transition
\cite{sch}.

A major step toward resolving the dilemma would be the determination of
an order parameter underlying the transition. The discovery of a
diverging length associated with the glass-like transition, similar
perhaps to the correlation length in a para-ferromagnetic transition
would enhance our understanding of glass-like transitions and of the
glassy state. From the structural point of view the glassy state is
characterized by some local (quenched) disorder and frustrated bonds
which inhibit the long range order and lead to the formation of
clusters of mesoscopic sizes. Thus, any theory which try to explain the
principal features of glass-like materials must take into account the
existence of such clusters. In this direction several models have been
proposed \cite{dom,coh,cha,dis,pal,vaz}.

During the past few years a new theoretical approach, the mode coupling
theory (MCT), has been developed which has motivated new experimental
investigations\cite{leu,ben,got}. This theory provides a generalized
kinetic equation approach to the density fluctuation dynamics in
supercooled fluids. It predicts two step relaxation processes ($\alpha$
and $\beta$ relaxations). The $\alpha$ relaxation process exhibits a
time scale which diverges at a crossover temperature $T_{c}$, which is
somewhat above the calorimetric glass transition temperature $T_{g}$,
and structural relaxation experiences rapid slowing down. On the other
hand the time scale of the $\beta$ process is considerably shorter,
does not diverge at $T_{c}$, and the $\beta$ process can therefore also
be observed below $T_{c}$. These predictions are directly accessible to
experimental verifications and have been tested for many glassforming
materials\cite{tao,li1,si1,li2,cu1,si2,cu2,wut,cu3,yos}.

Recently we have presented a fractal model which explains the
hysteresis loss effect observed in supercooled liquids \cite{vaz}. The
determinant influence of the fractal properties of the amorphous
structure in the system dynamics was emphasized introducing a fractal
distribution of cluster sizes. However, in that occasion, the nature
and origin of this fractal clusters was not analyzed.

In the present paper a cluster model for the relaxation in glass-like
materials is proposed following the general ideas presented in
\cite{vaz}.
As in previous works the relaxation of the system is obtained by
superposition of independent relaxing clusters with a size dependent
relaxation time. Nevertheless, in our model, differing from other
formulations, a cluster exists through a series of relaxation process
giving rise to a dynamically correlated cluster (DCC). In this way the
slow relaxation dynamics and other features of glass-like materials are
explained. Results are compared with experimental data reported in the
literature.

The paper is organized as follows. In section 2 the formation of DCC is
analyzed. The main ideas and concepts about DCC and cluster dynamics
are introduced. Then, the influence of the distribution of DCC on
relaxation dynamics is investigated. It is obtained that the relaxation
follows a two step process which may be identified with the $\alpha$
and $\beta$ processes. Comparison with experimental data and MCT
predictions is presented in section 3. Finally the conclusions are
given in section 4.

\section{Dynamic percolation model}

\subsection{DCC relaxation time}

Molecular dynamics simulations of soft-sphere mixtures \cite{miy} have
revealed that in addition to the stochastic jump motion of single atoms
correlated jump motions, in which a cluster of several atoms jump at
successive closed times by permuting their positions, take place.
Similar results have been obtained in molecular dynamics simulations of
water, a frustrated system with multiple random hydrogen bond network
structures \cite {sas}. Besides, MonteCarlo simulations of spin glasses
\cite{bin1,das} also revealed the existence of such cooperative
relaxation dynamics.

These results suggest that the cooperative motion of particles (atoms,
spins, etc.) in glass-like materials does not take place as the motion
of frozen clusters but as successive concatenations of single particle
motions. Hence, in this sense, a cluster is formed through the
perturbation propagation from one particle to its neighbors. This kind
of cluster will be referred here as a dynamically correlated cluster
(DCC). For short times only small clusters participate in the
relaxation dynamics however, with increasing time, larger and larger
clusters are incorporated.

Let us denote a characteristic physical quantity of the $i$-th DCC
particle by $x_{i}$. It may be the displacements around an equilibrium
position in ordinary glasses, the magnetic moment in a spin glass, or
other. $x_{i}$ will be random variable following certain distribution,
with zero mean value and variance $\langle x_i^2\rangle=\delta^2$,
$\delta$ being a typical value of fluctuations. Moreover, let
$\tau_{0}$ be a characteristic time scale of the fluctuations in
$x_{i}$.

The different DCC configurations will be then characterized by the sum
\begin{equation}
x=\sum_{i=1}^{s}x_{i}\ ,
\label{eq:1}
\end{equation}
where $s$ is the number of particles in the cluster, $1\geq s<\infty$.
In general the fluctuations $x_{i}$ are correlated but, in a first
approximation, they may be considered independent. In such an
approximation the central limit theorem\cite{fel} tells us that, for
large $s$, $P_s(x)$ the limit distribution of the sum in equation
(\ref{eq:1}) follows a Gaussian distribution with variance $\delta
s^{1/2}$, i.e.
\begin{equation}
P_s(x)=\frac{1}{\sqrt{2\delta^2s}}\exp\bigg(
-\frac{x^2}{2\delta^2s}\bigg)\ .
\label{eq:2}
\end{equation}

$P_s(x)$ represents the relative number of microstates or
configurations associated to a value of $x$, for a cluster of size $s$.
The entropy of this 'temporally isolated' subsystem is thus given by
\begin{equation}
S_s(x)=k\ln P_s(x)=-\frac{k}{2\delta^2s}x^2+\text{const.}\ ,
\label{eq:3}
\end{equation}
and the force associated to an entropy change
\begin{equation}
f_s(x)=T\frac{\partial S_s(x)}{\partial x}=-\frac{kT}{\delta^2s}x=-c x\
.
\label{eq:4}
\end{equation}

Now, from the kinetic point of view, the relaxation of the DCC may be
reduced to the Brownian motion of a harmonic oscillator with
generalized coordinate $x$, under the elastic force in equation
(\ref{eq:4}) and with diffusion coefficient $D=\delta^2/\tau_{0}$. In
this mean field approach the normalized relaxation is given
by\cite{ans}
\begin{equation}
\phi_s(t)=\frac{\langle x(t)\rangle}{\langle 
x(0)\rangle}=\exp\Big(-\frac{t}{\tau_s}\Big)\ ,
\label{eq:5}
\end{equation}
with the size dependent relaxation time
\begin{equation}
\tau_s=\frac{kT}{D}\frac{1}{c}=\tau_0 s\ .
\label{eq:6}
\end{equation}

Thus, in this mean field approximation the relaxation time of a DCC is
proportional to its size and a perturbation of the equilibrium state
relaxes exponentially towards equilibrium. As will be seen, the
existence of a broad distribution modifies the relaxation rate.

\subsection{Relaxation dynamics}

Let $n_s$ be the cluster numbers, i.e. the number of DCC with size $s$
per unit lattice. We assume that, in spite of the dynamic nature of the
clusters, the scaling assumption about the cluster number still holds.
Therefore, if the system is close to percolation
$\epsilon=(1-T/T_c)\sim 0$ the scaling assumption establishes that
\cite{sta}
\begin{equation}
n_s=s^{-\tau }f[|\epsilon| s^\sigma]\ ,
\label{eq:7}
\end{equation}
where $\tau$ and $\sigma$ are scaling exponents and $f(x)$ is a cutoff
function ($f(z\ll 1)\approx1$, and $f(z\gg 1)\ll 1$). Thus, the
averaged normalized decay will be given by
\begin{equation}
\phi (t)=\sum sn_s\phi_s(t)\Big/\sum sn_s\ . 
\label{eq:8}
\end{equation}

Close to percolation the sum is dominated by the contribution of large
cluster and we may substitute the sum by an integral. Thus, substituing
equations (\ref{eq:5}), (\ref{eq:6}) and (\ref{eq:7}) in equation
(\ref{eq:8}), with the variable change $z=t/\tau_s$, it is obtained
\begin{equation}
\phi(t)=\frac{\int_0^{t/\tau_0} dz
z^{a-1}e^{-z}f[(t/\tau_\epsilon)^\sigma z^{-\sigma}]}{\int_0^{t/\tau_0}
dz z^{a-1}f[(t/\tau_\epsilon)^\sigma z^{-\sigma}]}\ ,
\label{eq:9}
\end{equation}
where $\tau_\epsilon=\tau_0|\epsilon|^{-1/\sigma}$ and $a=\tau-2$

The first thing we note from this expression is the existence of two
characteristic times $\tau_0$ and $\tau_\epsilon$. The former is a
characteristic time for the microscopic motion while the last one is
the characteristic time of the largest clusters with size
$\sim\epsilon^{-1/\sigma}$.

It is tentative to associate these characteristic times with the
$\beta$ and $\alpha$ relaxation processes, respectively. Moreover, the
distribution of cluster sizes is divided in two regions. The initial
part which follows a power law decay and a second part with a sharp
cutoff. The cutoff size is of the order of $\epsilon^{-1/\sigma}$ with
characteristic time $\tau_\epsilon$. Hence, $\tau_\epsilon$ divides the
time scale in two regions where the relaxation function will show
different behaviors. These two time scales are analyzed below.

\subsubsection{$\beta$ relaxation}

For $t\ll\tau_\epsilon$ equation (\ref{eq:9}) is approximated by
\begin{equation}
\phi(t)\approx
a\Big(\frac{t}{\tau_0}\Big)^{-a}\gamma\Big(\frac{t}{\tau_0} ,a\Big)\ ,
\label{eq:10}
\end{equation}
where $\gamma(x,a)$ is the incomplete Gamma Euler function. Moreover,
for $t\ll\tau_0$, using the series expansion of $\gamma(x,a)$ for small
$x$, we obtain
\begin{equation}
\phi(t)\approx 1-\frac{a}{1+a}\frac{t}{\tau_0}\approx\exp\Big(
-\frac{a}{1+a}\frac{t}{\tau_0}\Big)\ ,
\label{eq:11}
\end{equation}
while for $t\gg\tau_0$, $\gamma(t/\tau_0,a)\approx\Gamma(a)$, resulting
\begin{equation}
\phi(t)\approx\Gamma(1+a)\Big(\frac{t}{\tau_0}\Big)^{-a}\ .
\label{eq:11a}
\end{equation}
Thus, the initial decay, due to single particle motions, is
exponential. However, for $t\gg\tau_0$ cooperative dynamics leads to
the formation of DCC, with power law distribution sizes, giving the
power tail in equation (\ref{eq:11a}). The fractal behavior in the
distribution of DCC sizes leads to fractal properties in the time
decay.

The imaginary part of the susceptibility $\chi{''}$ is related to the
relaxation function through the equation
\begin{equation}
\chi^{''}(\omega)=\omega\text{FT}^{'}[\phi(t)]\ ,
\label{eq:12}
\end{equation}
where $\text{FT}^{'}$ denotes the real part of the Fourier transform.
Then, for $\tau_0\ll t\ll\tau_\epsilon$, using equation (\ref{eq:11a})
it is obtained
\begin{equation}
\chi^{''}(\omega)\approx a\sec\Big(a\frac{\pi}{2}\Big)(\omega\tau_0)^a\ .
\label{eq:13}
\end{equation}
This frequency dependence corresponds with the $\beta$ relaxation. The
characteristic time for these time scale is $\tau_0$ which does not
depend on $\epsilon$. Hence, if we assume that the divergences near
$T_c$ are given by the percolative nature of the DCC, $\tau_0$ will not
diverge at $T=T_c$ as it is expected for the $\beta$ process.

\subsubsection{$\alpha$ relaxation}
Now let us investigate the influence of the cutoff in the distribution
of cluster sizes in the long time ($t\gg\tau_\epsilon$) relaxation
dynamics. In this case the form of the cutoff function $f(x)$ is
determinant and cannot be rulled out. We will assume that this cutoff
function is of the form
\begin{equation}
f(x)=\exp(-x^{\theta_\pm})\ ,
\label{eq:14}
\end{equation}
where $\theta_-$ and $\theta_+$ stand for $\epsilon<0$ and
$\epsilon>0$, respectively.

An asymptotic expansion, for $t\gg\tau_\epsilon$, of $\phi(t)$ defined
in equation (\ref{eq:9}) with $f(x)$ in equation (\ref{eq:14}) give the
estimate
\begin{equation}
\phi(t)\sim\exp\Big[-\Big(\frac{t}{\tau_{\epsilon\pm}}\Big)^{\beta_\pm}\Big]
\ .
\label{eq:17}
\end{equation}
with
\begin{equation}
\tau_{\epsilon\pm}=c_{\sigma,\theta_\pm}\tau_\epsilon\ ,
\label{eq:15}
\end{equation}
\begin{equation}
\beta_\pm=\frac{\sigma\theta_\pm}{1+\sigma\theta_\pm}\ .
\label{eq:16}
\end{equation}
$c_{\sigma,\theta_\pm}$ is a constant which does not depend on
$\epsilon$ and, therefore, $\tau_{\epsilon\pm}$ diverges as
$\tau_\epsilon$ near $T_c$.

Equation (\ref{eq:17}) is the well known Kolraush stretched exponential
often used to fit the experimental data in the $\alpha$ relaxation
process.  Moreover, the characteristic time
$\tau_{\epsilon\pm}\propto\tau_\epsilon$ diverges at $T=T_c$ as it is
expected for the $\alpha$ relaxation process.

\subsubsection{Static viscosity}

The static viscosity is related to the mean cluster size through the
expression
\begin{equation}
\eta=G\langle\tau_s\rangle=G\tau_0\langle s\rangle\ ,
\label{eq:18}
\end{equation}
where $G$ is an elasticity modulus and $\langle s\rangle$ is given by
\begin{equation}
\langle s\rangle=\sum s^2n_s\Big/\sum sn_s\ . 
\label{eq:19}
\end{equation}

For $T\stackrel{>}{\sim}T_c$ the mean cluster size defined in equation
(\ref{eq:19}) will diverge according to $\langle
s\rangle\sim|\epsilon|^{-1/\sigma}$\cite{sta} and, therefore,
\begin{equation}
\eta\sim\tau_\epsilon\sim\tau_{\epsilon\pm}\sim\langle 
s\rangle\sim|\epsilon|^{-1/\sigma}\ .
\label{eq:20}
\end{equation}

Hence, $\tau_{\epsilon\pm}$ the characteristic time for the $\alpha$
relaxation process shows the same temperature dependence of the static
viscosity.

\section{Discussion}

The present model exhibits the two step relaxation predicted by the
MCT\cite{leu,ben,got}. The exponent $a$ of the $\beta$ process
($\chi^{''}\sim\omega^a$) is identified here with $a=\tau-2$. Moreover,
since $2\leq\tau\leq 5/2$ as it is obtained in conventional
percolation\cite{sta} then $0\leq a\leq1/2$, which is in the range
predicted by the MCT. On the other hand, the Kolraush stretched
exponent of the $\alpha$ process, given by equation (\ref{eq:16}),
satisfies $0\leq\beta_\pm<1$ which is also the range predicted by the
MCT. Moreover, the characteristic time of the $\alpha$ process has the
same temperature dependence as the static viscosity, in agreement with
the MCT.

Thus, our model seems to arrive to the same results obtained by the
MCT, which have been extensively tested in light scattering and neutron
scattering
experiments\cite{tao,bor,li1,si1,li2,cu1,si2,cu2,wut,cu3,yos}. In order
to perform a more accurate comparison let us investigate the behavior
of $\chi^{''}$ around the minimum between the $\alpha$ peak and the
$\beta$ process. This minimum should be near the frequency
$1/\tau_\epsilon$ since it marks the transition from the $\beta$ to the
$\alpha$ process in our model. In this frequency range we will have the
contribution of the $\beta$ process with relax according to
$(t/\tau_0)^{-a}$ and of the $\alpha$ process relaxing as
$\exp[-(t/\tau_{\epsilon\pm})^{\beta_\pm}]\approx
1-(t/\tau_{\epsilon\pm})^{\beta_\pm}$. Thus, a gross estimate of the
relaxation function in the time scale will be given by considering both
contributions which, resulting
\begin{equation}
\phi(t)\sim|\epsilon|^{a/\sigma}\Big[\Big(\frac{t}{
\tau_{\epsilon\pm}}\Big)^{-a}-\Big(\frac{t}{\tau_{
\epsilon\pm}}\Big)^{b}\Big]\ ,
\label{eq:21}
\end{equation}
where $b=\beta_\pm-a$. Substituting this expression in equation
(\ref{eq:12}) and assuming $a,b\ll 1$ it is obtained
\begin{equation}
\chi^{''}(\omega)\sim|\epsilon|^{a/\sigma}
\Big[a(\omega\tau_{\epsilon\pm})^a+b(\omega\tau_{\epsilon\pm})^{-b}\Big]\ ,
\label{eq:22}
\end{equation}
This expression constitutes an interpolation formula often used to fit
the experimental data. The values of $a$ and $b$ obtained from the fit
to experimental data are in general small, thus our supposition $a,b\ll
1$ is consistent with experiments. From this expression we obtain that
the position of the minimum $\omega_{\text{min}}$ and the value of
$\chi^{''}(\omega)$ at the minimum $\chi^{''}_{\text{min}}$ satisfies
\begin{equation}
\omega_{\text{min}}\sim|\epsilon|^{1/\sigma}\ ,
\label{eq:23}
\end{equation}
\begin{equation}
\chi^{''}_{\text{min}}\sim|\epsilon|^{a/\sigma}\ .
\label{eq:24}
\end{equation}

If $a/\sigma=1/2$ then we obtain the MCT prediction which is consistent
with the light scattering data\cite{li1,li2,cu2,wut,cu3,yos}. However,
under this assumption the static viscosity in equation (\ref{eq:20})
will diverge near $T_c$ according to $\eta\sim|\epsilon|^{-\gamma}$,
with $\gamma=1/2a$, while the MCT predicts $\gamma=1/(2a)+1/(2b)$.
Thus, when $a/\sigma=1/2$ we obtain the MCT predictions, except for the
exponent $\gamma$. Our result is more general since the ratio
$a/\sigma$ is not necessarily restricted to this value.

The value of $a$ obtained from the fit to light scattering data for
different glass-forming materials is found temperature independent and
around $0.3$. Therefore, $\tau=2+a\sim 2.3$ is slightly larger than the
one obtained in percolation theory in three dimensions $\tau\approx
2.15$\cite{st1}. Besides, as was shown above, $a/\sigma=1/2$ in order
to obtain a good agreemen with the MCT and experiments. In the present
model $a/\sigma=(\tau-2)/\sigma$ is the critical exponent $\beta$ of
conventional percolation\cite{sta} which is also slightly smaller
($0.4$) in three dimensions\cite{st1}. These results suggest that the
scaling exponent $\tau$ is larger than the one obtained in conventional
percolation theory. The scaling assumption holds but the power decay is
stronger.

On the other hand, the MCT predicts that the exponent $\beta$ of the
$\alpha$-process does not depend on temperature. However, it is found
to be temperature dependent\cite{li1,li2,bor}. It takes a constant
value above $T_c$ and then decreases near $T_c$, taking again a
constant value below $T_c$. This high and low temperature limits are
identified in our model with $\beta_-$ and $\beta_+$, respectively. The
main contribution to this temperature dependence is given by
$\theta_\pm$ in equation (\ref{eq:16}). For instance, in conventional
percolation theory\cite{sta} $\sigma\theta_-=1$ ($\beta_-=1/2$) and
$\sigma\theta_+=1-1/d$ ($\beta_+=(d-1)/(2d-1)$), where $d$ is the space
dimensionality, and therefore $\beta_->\beta_+$. While this behavior is
in agreement with the experimental observations the values of
$\beta_\pm$, using conventional percolation exponents, are smaller than
those observed in experiments. Thus, we found again that conventional
percolation theory can not give a precise quantitative description of
the properties of glass-like materials. However, the scaling
assumptions are still valid, but with different scaling exponents. To
obtain a good agreement with the experimental observations
$\sigma\theta_\pm$ should take larger values which implies that, in
glass-forming materials, the cutoff for larger DCC sizes is stronger
than the one expected in conventional percolation.

This sharp cutoff for large sizes may be attributed to frustration
effects which destroy the long range order. The existence of
frustration, which is an inherent property of glass-like materials, may
lead to different critical exponents. The percolation problem including
frustration effects (frustrated percolation) is a subject of recent
study, and exhibits properties which are different from the
corresponding unfrustrated case\cite{con}.

Finally we want to mention that other authors have used the idea of
DCC, for instance the models by Domb {\em et al}\cite{dom}, Cohen {\em
et al}\cite{coh}, and Chamberlin\cite{cha}. However, these models does
not analyze the properties of glass-like systems around the dynamic
transition, which is the main contribution of the present work.

\section{Conclusions}
The complex dynamics of glass-forming materials was reduced to the
formation of independent dynamic correlated clusters (DCC) with a size
dependent relaxation time. The relaxation of the system was obtained as
a superposition of exponential relaxations over the distribution of the
DCC sizes.  In this way we have obtained the two step relaxation
process observed in supercooled liquids.

Our model contains as a particular case the MCT predictions, it is more
simple and it is based in a very general principle, the scaling
assumption for the distribution of DCC near $T_c$. Besides, the present
model gives the correct temperature dependence for the Kolraush
exponent $\beta$, while the MCT predicts a constant value in
disagreement with the experimental bahavior.

The comparison of our results with the experimental data for
glass-materials suggest that the percolation critical exponent $\tau$
should be larger than the one obtained in conventional percolation
theory. Moreover, the cutoff for larger DCC sizes should also be
stronger. These results were attributted to the existence of
frustration effects, which inhibit the formation of long range order
dynamic structures.

We conclude that the use of geometrical descriptions, the distribution
of DCC size for instance, constitute an alternative approach to
describe the dynamics of glass-like materials. The properties of these
systems near the dynamic transition can be predicted using conventional
tools of standard critical phenomena such as the scaling hypothesis.

\end{multicols}


\begin{thebibliography}{99}

\bibitem{sch}  G. W. Scherrer, J. Non-Cryst. Solids {\bf 123}, 75 (1990).

\bibitem{dom}  C. Domb, A. A. Maradudin, E. W. Montroll, and G. H. Weiss,
Phys. Rev. {\bf 115}, 24 (1959).

\bibitem{coh}  M. H. Cohen and G. Grest, Phys. Rev. B {\bf 24}, 4991 (1981).

\bibitem{cha}  R. W. Chamberlin, J. Appl. Phys. {\bf 76}, 6401 (1994).

\bibitem{dis} L. A. Dissado and R. M. Hill, Proc. Roy. Soc. A {\bf
390}, 131 (1983).

\bibitem{pal}  R. G. Palmer, D. L. Stein, E. Abrahams, and P. W. Anderson,
Phys. Rev. Lett. {\bf 53}, 958 (1984).

\bibitem{vaz}  A. V\'{a}zquez and O. Sotolongo-Costa, J. Chem. Phys. 
{\bf 106}, 3772 (1997).

\bibitem{leu} E. Leutheusser, Phys. Rev. A {\bf 29}, 2765 (1984).

\bibitem{ben} U. bengtzelius, W. G\"otze, and L. Sjolander, J. Phys. C 
{\bf 17}, 5915 (1984).

\bibitem{got} W. G\"otze, Z. Phys. B {\bf 56}, 139 (1984); W. G.
G\"otze and L. Sj\"ogren, J. Phys. C {\bf 21}, 3407 (1988).

\bibitem{tao} N. J. Tao, G. Li, nad H. Z. cummins, Phys. Rev. Lett. 
{\bf 66}, 1334 (1991).

\bibitem{bor} L. B\"orjesson, M. Elmroth, and L. M. Torell, J.
Non-Cryst. Sol. {\bf 131-133}, 139 (1991).

\bibitem{li1} G. Li, W. M. Du, X. K. Chen, and H. Z. Cummins, Phys.
Rev. A {\bf 45}, 3867 (1992).

\bibitem{si1} D. L. Sidebottom, R. bergan, L. B\''orjesson, and L. M. 
Torell, Phys. Rev. Lett. {\bf 68}, 3587 (1992).

\bibitem{li2} G. Li, W. M. Du, A. Sakai, nad H. Z. Cummins, Phys. Rev.
A {\bf 46}, 3343 (1992).

\bibitem{cu1} H. Z. cummins, W. M. Du, M. Funchs, W. G\''otze, S. 
Hildebrand, A. Latz, G. Li, and N. J. tao, Phys. Rev. E {\bf 47}, 4223 (1993).

\bibitem{si2} D. sidebottom, R. Bergman, L. B\"orjesson, and L. M.
Torell, Phys. rev. Lett. {\bf 71}, 2260 (1993).

\bibitem{cu2} H. Z. cummins, W. M. Du, F. Funchs, W. G\"otze, A. Latz,
and N. J. tao, Physica A {\bf 201}, 207 (1993). 

\bibitem{wut} J. Wutte, J. Hernandez, G. Li, G. Goddens, H. Z. Cummins,
F. Fujara, W. Petry, and H. Sillescu, Phys. Rev. Lett. {\bf 72}, 3052 (1994).

\bibitem{cu3} H. Z. Cummins, G. Li, W. Du, Y. H. Hwang, and G. Q. Shen,
Prog. Theor. Phys. No. {\bf 126}, 21 (1997).

\bibitem{yos} A. Yoshihara, H. Sato, and S. Kojima, Prog. Theor. Phys. 
{\bf 126}, 423 (1997).

\bibitem{miy}  H. Miyagawa, Y. Hiwatari, B. Bernu, and J. P. Hansen, J.
Chem. Phys. {\bf 88}, 3879 (1988).

\bibitem{sas}  M. Sasai, I. Ohmine, and R. Ramaswamy, J. Chem. Phys. 
{\bf 96}, 2045 (1992).

\bibitem{bin1}  K. Binder and K. Schr\"{o}der, Phys. Rev. B {\bf 14}, 
2142 (1976).

\bibitem{das}  C. Dasgupta, S. Ma, and C. Hu, Phys. Rev. B {\bf 20}, 
3837 (1979).

\bibitem{fel} W. Feller, {\em An Introduction to Probability Theory 
and its Applications}, vol. I (Wiley, New York, 1966).

\bibitem{ans} A. Anselm, {\em Osnovi Statisticheskoi Fisiki i 
Thermodinamiki} (Nauka, Moscu, 1973).

\bibitem{sta}  D. Stauffer, Phys. Rep. {\bf 54}, 1 (1979).

\bibitem{wei} R. Weiler, S. S. Blaser, and P. B. Macedo, J. Chem. 
Phys. {\bf 73}, 4147 (1969).

\bibitem{mac} P. B. Macedo and A. Napolitano, J. Chem. Phys. {\bf 49}, 
1887 (1968).

\bibitem{st1} Reference \cite{sta}, table 2 page 25.

\bibitem{con} A. Coniglio, Prog. Theor. Phys. {\bf 126}, 281 (1997); 
and references therein.

\end{thebibliography}
\end{document}